\documentclass[iop]{emulateapj}

\def\mybf{}

\def\sqig{$\sim$}
\def\sun{$_\odot$}

\def\Swift{{\it Swift}}
\def\RXTE{{\it RXTE}}
\def\INTEGRAL{{\it INTEGRAL}}
\def\XMM{{\it XMM-Newton}}

\def\2s{2S\,0114+650}
\begin{document}
\title{
Superorbital Periodic Modulation in Wind-Accretion High-Mass
X-ray Binaries from \Swift\ BAT Observations}

\author{Robin H.~D. Corbet\altaffilmark{1,2}}
\author{Hans A. Krimm\altaffilmark{3,4}}

\altaffiltext{1}{University of Maryland, Baltimore
County, MD, USA; corbet@umbc.edu}

\altaffiltext{2}
{CRESST/Mail Code 662, X-ray Astrophysics Laboratory,
NASA Goddard Space Flight Center, Greenbelt, MD 20771, USA}

\altaffiltext{3}
{Universities Space Research Association,
10211 Wincopin Circle, Suite 500, Columbia, MD 21044, USA}

\altaffiltext{4}
{CRESST/Mail Code 661, Astroparticle Physics Laboratory,
NASA Goddard Space Flight Center, Greenbelt, MD 20771, USA}

\begin{abstract}
We report the discovery using data from the \Swift\ Burst Alert Telescope (BAT) 
of superorbital modulation in the wind-accretion supergiant high-mass X-ray binaries
4U 1909+07 (= X 1908+075), IGR J16418-4532, and IGR J16479-4514.
Together with already known superorbital periodicities
in \2s\ and IGR J16493-4348, the systems exhibit a monotonic relationship between
superorbital and orbital periods. These systems include both supergiant fast X-ray
transients (SFXTs) and classical supergiant systems, and have a range of inclination
angles. This suggests an underlying physical mechanism which is connected to the orbital
period. 
In addition to these sources with clear detections of superorbital periods, 
IGR J16393-4643 (= AX J16390.4-4642) is identified as a system that may have superorbital
modulation due to the coincidence of low-amplitude peaks in power spectra derived from BAT, RXTE PCA,
and INTEGRAL light curves. 1E 1145.1-6141 may also be worthy of further attention due to
the amount of low-frequency modulation of its light curve. 
However, we find that the presence of superorbital modulation is not a universal feature
of wind-accretion supergiant X-ray binaries.
\end{abstract}
\keywords{stars: individual (2S 0114+650, 1E 1145.1-6141, IGR J16393-4643, IGR J16418-4532, IGR J16479-4514, 
IGR J16493-4348, 4U 1909+07) --- stars: neutron --- X-rays: stars}

\section{Introduction}

Superorbital modulation is seen in a variety of 
X-ray binaries. A review of superorbital modulation in several types of systems is presented
by \citet{Kotze2012}.
In some cases such as Her X-1, SMC X-1 and
LMC X-4, where accretion occurs by Roche-lobe overflow via
an accretion disk onto a neutron star, the mechanism driving superorbital
modulation can be understood as either precession of the accretion
disk \citep[e.g.][]{Petterson1975} or of the neutron star \citep[e.g.][]{Postnov2013}. Irradiation of the accretion disk by
the central X-ray source provides a possible mechanism for 
driving disk precession \citep[e.g.][and references therein]{Ogilvie2001}.
Be star systems also exhibit long timescale, possibly periodic,
variability at optical wavelengths. This long timescale variability has been claimed
to be correlated with orbital period \citep{Raj2011}.

A more puzzling variety of superorbital variability was found in
a supergiant high-mass X-ray binary (sgHMXB). 
The sgHMXBs can be broadly classified into ``classical''
systems, which may suffer from high levels of absorption,
and supergiant fast X-ray transients \citep[SFXTs; e.g.][]{Blay2012,Sidoli2013}.
In the sgHMXB \2s\
there are three
periodicities:
a \sqig9700 s neutron star rotation
period, an 11.6 day orbital period, and a 30.7 day superorbital modulation \citep{Corbet1999,Wen2006,Farrell2008}.
A question has been whether \2s\ is exceptional, perhaps because of
its unusually long pulse period, or whether other wind-accretion
HMXBs also show similar superorbital periodicities. If similar
behavior is found in other systems, then this may provide
a way to determine, or at least constrain, the underlying mechanism.
A suggestion that the phenomenon might be more general than just the case of \2s\ came
when a superorbital period was found in the sgHMXB IGR J16493-4348 \citep{Corbet2010b}.

The \Swift\ BAT provides an excellent way to monitor sgHMXBs.
These systems are often highly absorbed, which presents difficulties
for an instrument such as the Rossi X-ray Timing Explorer (\RXTE) All Sky Monitor (ASM) which is sensitive in the 2 - 12 keV
band \citep{Levine1996}. The \Swift\ BAT's sensitivity to higher energy X-rays ($>$ 15 keV) provides a way to peer
through this absorption. We present here a review of our searches of BAT light curves
of sources thought to be HMXBs, in order to find additional sources that
may also display superorbital modulation. In a few cases we also employ data
collected from Galactic plane scans \citep{Markwardt2006} made with the \RXTE\ Proportional Counter Array (PCA). 
The large effective area
of the PCA enables observations to be made in the lower energy
range of 2 - 10 keV. Although the PCA data cover only a limited fraction
of the sky they have greater sensitivity than the \RXTE\ ASM.
MAXI light curves \citep{Sugizaki2011} are not available for the majority of systems considered here.

We present here the results of a search for superorbital modulation in
additional wind-accretion supergiant HMXBs. We find three new systems with clear superorbital
modulation, initial reports of which were made in \citet{Corbet2013a,Corbet2013b}.
We also find hints of modulation in two other systems.
Although the number of systems is small, three new systems plus two previously known,
we note a monotonic relationship
between orbital and superorbital periods. We consider possible mechanisms that
might cause superorbital modulation in some, but not all, sgHMXBs.

\section{Data and Analysis}

The Burst Alert Telescope (BAT) on board the \Swift\ satellite \citep{Gehrels2004} is
described in detail by \citet{Barthelmy2005}. It uses a 2.7 m$^2$ coded-aperture mask 
and a 0.52 m$^2$  CdZnTe detector array. The BAT has a wide field of view,
1.4 sr half-coded, 2.85 sr 0\% coded. The pointing direction of \Swift\ is driven by
the narrow-field XRT and UVOT instruments on board the satellite.
The BAT typically observes 50\%--80\% of the sky each day.
We used data from the \Swift/BAT transient monitor \citep{Krimm2006,Krimm2013}
covering the energy range 15 - 50 keV, and selected data with time resolution
of \Swift\ pointing durations.
The transient monitor data are available shortly after observations have been
performed. 
The light curves considered here cover the time range of MJD 53,416 to 56,452 (2005-02-15 to 2013-06-09).
The light curves of some sources, not including the ones discussed in detail here,
were more recently added to the
analysis and hence have shorter durations.
BAT light curves are also available from the catalogs such as described by \citet{Tueller2010}.
However, the most recent BAT catalog is from 70 months of data \citep{Baumgartner2013}
and the transient monitor light curves are hence of longer duration.
The transient monitor light curves generally cover more than 3000 days, approximately 50\% longer 
than the 70-month catalog light curves.

We used only data for which the data quality flag (``DATA\_FLAG'') was
set to 0, indicating good quality. In addition, we found that even data flagged
as ``good'' {\mybf were} sometimes suspect. In particular we identified a small number of data points with 
very low fluxes and implausibly small uncertainties. We therefore removed
these points from the light curves.
A total of 1244 light curves were available, this includes 106 blank fields that are
used for test purposes. 

To search for periodic modulation in the light curves, we calculated
discrete Fourier transforms (DFTs) of all available light curves. 
We calculated the DFTs for a frequency range which corresponds to
periods of between 0.07 days to the length of the light curves -
i.e. generally \sqig3000 days.
The contribution of each
data point to the power spectrum was weighted by its uncertainty
using the ``semi-weighting'' technique \citep{Corbet2007a,Corbet2007b}. This takes into account
both the error bars on each data point and the excess variability of the
light curve. \citet{Scargle1989} notes that the weighting of data points in a power spectrum
can be compared to combining individual data points.
In this way, the use of semi-weighting is
analogous to combining data points using the semi-weighted mean \citep{Cochran1937,Cochran1954}.
We oversampled the DFTs by a factor of five compared to their nominal resolution.
Calculations of the significance of peaks seen are expressed in terms of false alarm probability
\citep[FAP; ][]{Scargle1982} which takes into account the DFT oversampling.
Uncertainties in periods are generally derived using the expression
of \citet{Horne1986}.
In the figures showing power spectra we mark in ``white noise'' 99.9\% and
99.99\% significance levels. However, many sources exhibit noise continua which
are not ``white''. In our calculations of FAP, we therefore determined local
noise levels by fitting the continuum power levels in a narrow frequency range around each
peak of interest.

\section{Sources with Previously Reported Periodic Superorbital Modulation}

\subsection{\2s}
\2s\ is an unusual HMXB system that has
an exceptionally long pulse period of \sqig 9700 s \citep[e.g.][]{Corbet1999}.
There has been controversy over the spectral classification of the mass donor, but
\citet{Reig1996} derive a spectral type of B1 Ia.
From optical radial velocity measurements, \citet{Grundstrom2007} determine
an orbital period of 11.5983 $\pm$ 0.0006 days and a moderate eccentricity of 0.18 $\pm$ 0.05.
The orbital period is also seen in the RXTE ASM light curve \citep{Corbet1999,Wen2006}.
A 30.7 $\pm$ 0.1 day superorbital period was found by \citet{Farrell2006}
from \RXTE\ ASM observations, and the period was later refined
to 30.75 $\pm$ 0.03 days by \citet{Wen2006}.
\citet{Farrell2008} performed extensive \RXTE\ PCA observations
covering approximately 2 cycles of the superorbital period.
Although \citet{Farrell2008} found variations in the X-ray absorption on
the orbital period, they found no such changes over the superorbital period.
However, a significant increase in the photon index of the power-law model
used to fit the X-ray spectrum was reported at the minimum flux phase
of the superorbital period. \citet{Farrell2008} concluded that the superorbital
modulation was due to mass-accretion rate variations, although the mechanism
causing this could not be determined.

\begin{figure}
\epsscale{0.8}
\includegraphics[angle=-90,scale=0.3]{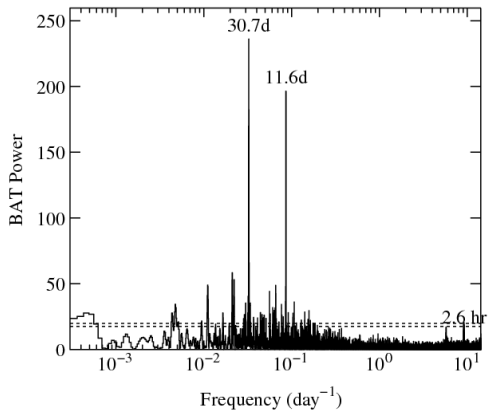}
\caption{Power spectrum of the BAT light curve of \2s.
Note that the superorbital peak at 30.7 days is stronger than
the orbital modulation at 11.6 days.
The horizontal dashed lines indicate ``white noise'' 99.9\% and
99.99\% significance levels.
}
\label{fig:0114_power}
\end{figure}

\begin{figure}
\epsscale{0.8}
\plotone{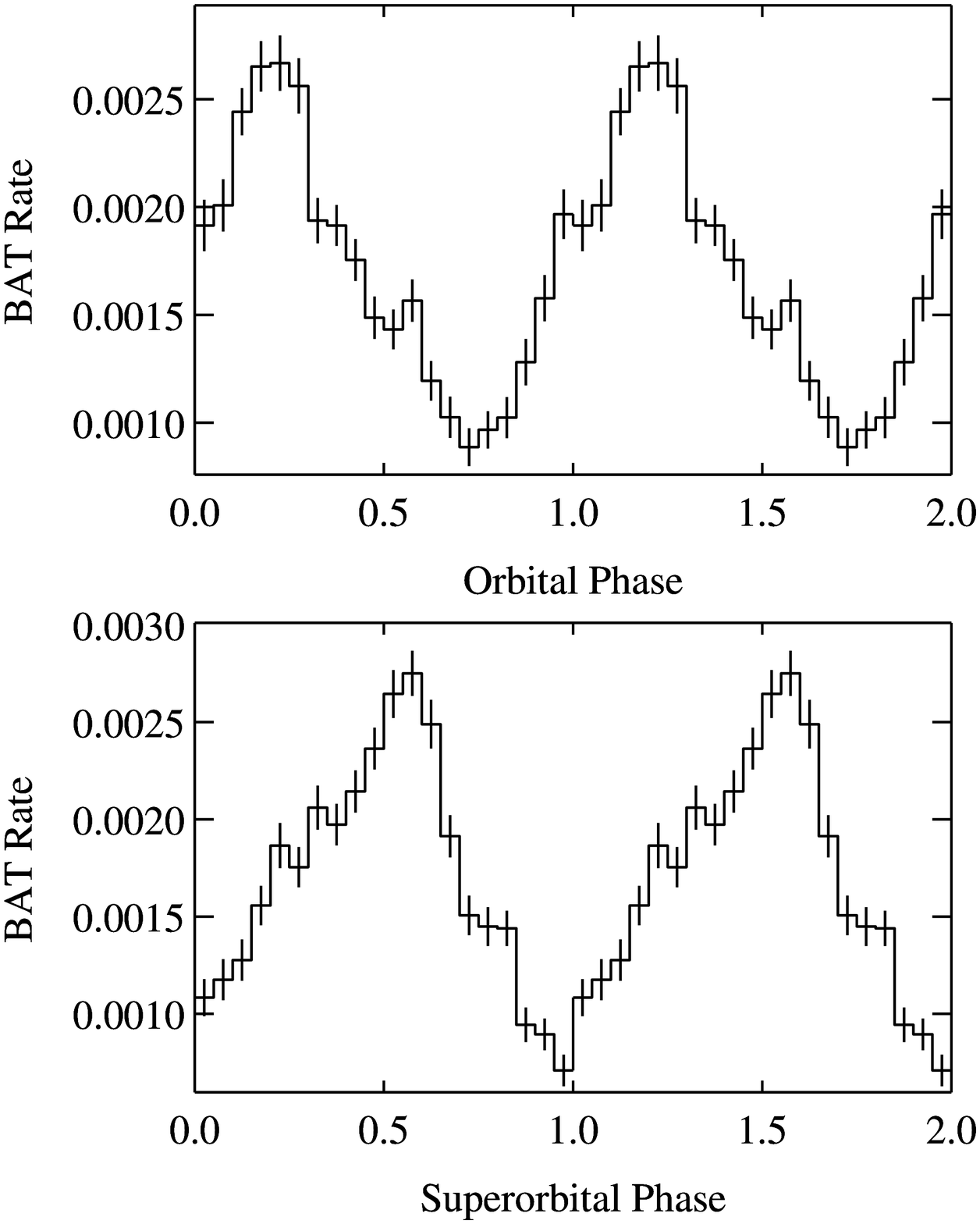}
\caption{Swift BAT light curve of \2s\ folded on its orbital period {\mybf (top)}
and folded on its superorbital period {\mybf (bottom)}. The period values are given
in Table \ref{table:sources}.
Phase zero for the orbital period is the time of periastron passage
from \citet{Grundstrom2007} and is MJD 51,824.8.
Phase zero for the superorbital period is the time of minimum flux from
\citet{Farrell2008} and is MJD 53,488. {\mybf We note that this 
definition of phase zero differs from the other sources
considered in this paper where phase zero for the superorbital modulation
is defined as the time of maximum flux.}
}
\label{fig:0114_fold}
\end{figure}

The power spectrum of the BAT light curve of \2s\ is shown in Figure \ref{fig:0114_power}, both the orbital
and superorbital periods are strongly detected, with the superorbital period
being stronger than the orbital modulation. We determine orbital and superorbital
periods of 11.591  $\pm$ 0.003 and 30.76 $\pm$ 0.03 days respectively.
The BAT light curve of \2s\ folded on the orbital and superorbital periods is shown
in Figure \ref{fig:0114_fold}. 
{\mybf For consistency with the work of \citet{Farrell2008} the light curve
folded on the superorbital period uses a definition of phase zero as the
time of minimum flux. However, for the other sources considered in this paper
we use the time of maximum flux as phase zero.}
Both the orbital and superorbital modulations are quasi-sinusoidal
and no evidence for an eclipse is seen in the light curve folded on the orbital period.

\subsection{IGR J16493-4348}
IGR J16493-4348 was discovered by \citet{Grebenev2005} 
and subsequent X-ray observations suggested that the source is an X-ray binary
\citep{Hill2008}. \citet{Nespoli2010} classified the infrared counterpart as a B0.5 I supergiant. 
A 6.8 day orbital period was independently found by \citet{Corbet2010b} and \citet{Cusumano2010}
using BAT 54 month survey data with the two groups finding periods of  6.7906 $\pm$ 0.0020 and 
6.782 $\pm$ 0.005 days respectively.
The BAT modulation was interpreted by \citet{Cusumano2010} as showing the presence
of an eclipse. \citet{Corbet2010b} confirmed the orbital period using PCA Galactic plane scan
data which gave an orbital period of 6.7851 $\pm$ 0.0016 days. 
In addition, \citet{Corbet2010b} noted the presence of a 20.07 $\pm$ 0.02 day superorbital period in the BAT
data which was confirmed by modulation at 20.09 $\pm$ 0.02 days in the PCA observations.
Pointed \RXTE\ PCA observations revealed a \sqig 1093 s pulse period \citep{Corbet2010c},
and pulse timing with the PCA yielded a mass function of 14.0 $\pm$ 2.3 M\sun\ \citep{Pearlman2013}
which confirms the interpretation of IGR J16493-4348 as a supergiant HMXB.

\begin{figure}
\epsscale{0.8}
\includegraphics[angle=-90,scale=0.3]{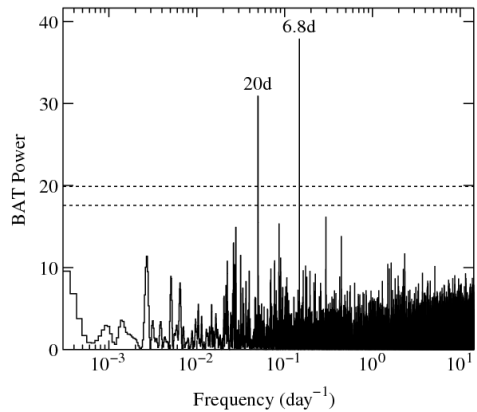}
\caption{Power spectrum  of the BAT light curve of IGR J16493-4348.}
\label{fig:16493_power}
\end{figure}

\begin{figure}
\epsscale{0.8}
\plotone{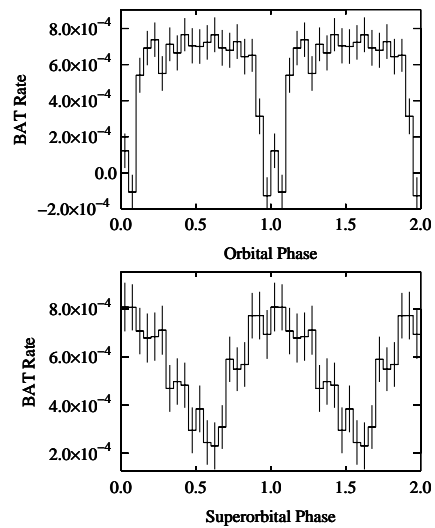}
\caption{Swift BAT light curve of IGR J16493-4348 folded on its orbital period {\mybf (top)}
and folded on its superorbital period {\mybf (bottom)}.
Phase zero for the orbital light curve corresponds to the center of the eclipse
and is MJD 54,175.92 \citep{Cusumano2010}.
Phase zero for the superorbital light curve corresponds to the time of maximum flux
and is MJD 54,265.1 \citep{Corbet2010b}.
}
\label{fig:16493_fold}
\end{figure}

The power spectrum of the BAT light curve of IGR J16493-4348 is shown in Figure \ref{fig:16493_power}.
This clearly shows the presence of the already known orbital and superorbital
periods. However, the statistical significances of the periods are somewhat less than previously
found from the BAT 54-month catalog data and the FAPs were \sqig10$^{-6}$ and 0.04 respectively.
We refine the period measurements to be 6.782 $\pm$ 0.001 and 20.07 $\pm$ 0.01 days
for the orbital and superorbital periods respectively.
The BAT light curve of IGR J16493-4348 folded on the orbital and superorbital periods is shown
in Figure \ref{fig:16493_fold}. The orbital modulation shows the presence of an
eclipse, while the superorbital modulation is quasi-sinusoidal.


\section{Sources with New Detections of Periodic Superorbital Modulation}

\subsection{IGR J16418-4532} \label{section:16418}

\citet{Chaty2008} determined that the optical counterpart of IGR J16418-4532 is probably an OB supergiant.
\citet{Rahoui2008} fitted the spectral energy distribution
of the likely 2MASS counterpart and found that this was
consistent with an O/B massive star classification {\mybf with a best fit spectral
type of O8.5, although the luminosity type could not be determined}.
IGR J16418-4532 exhibits large flux variability, classifying it as an SFXT {\mybf \citep{Romano2011,Romano2012,Sidoli2012}}.
Pulsations from the source were discovered
by \citet{Walter2006} and refined to a period of 1212 $\pm$ 6 s by \citet{Sidoli2012}.
A 3.74 day orbital period has been found for IGR J16418-4532
from \RXTE\ ASM and \Swift\ BAT observations
\citep[e.g.][]{Corbet2006,Levine2011}.
\INTEGRAL\ and \XMM\ observations of IGR J16418-4532 are discussed by \citet{Drave2013a}.

\begin{figure}
\epsscale{0.8}
\includegraphics[angle=-90,scale=0.3]{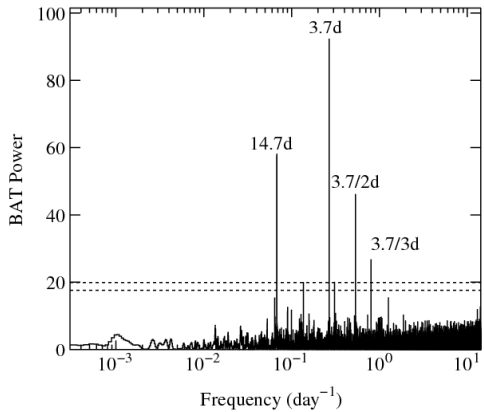}
\caption{Power spectrum of the BAT light curve of IGR J16418-4532.}
\label{fig:16418_power}
\end{figure}

\begin{figure}
\epsscale{0.8}
\plotone{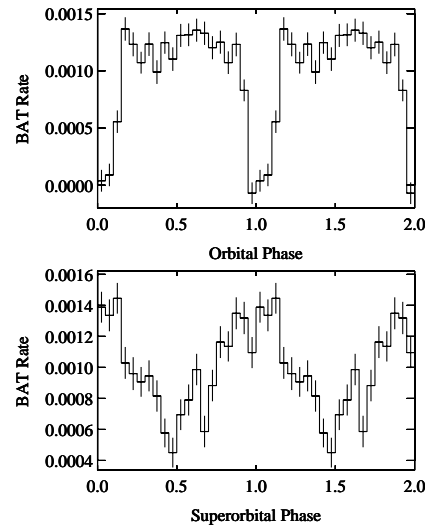}
\caption{Swift BAT light curve of IGR J16418-4532 folded on its orbital period {\mybf (top)},
and folded on its superorbital period {\mybf (bottom)}.
The period values are given
in Table \ref{table:sources}.
Phase zero for the orbital period corresponds to the time of minimum flux
and is MJD 52,735.84 \citep{Levine2011}.
For
the superorbital period phase zero corresponds to the time of maximum
flux and is MJD 55,994.6 (Section \ref{section:16418}).
}
\label{fig:16418_fold}
\end{figure}

The power spectrum of the BAT light curve of IGR J16418-4532 (Figure \ref{fig:16418_power}) shows
modulation at the 3.74 day orbital
period and the second and third harmonics of this.
In addition the power
spectrum  shows a peak near 14.7 days with an FAP of
$<$ 10$^{-6}$. The light curve folded on this period (Figure \ref{fig:16418_fold}) shows an approximately
sinusoidal modulation. From a sine wave fit to the light curve we
obtain:
\begin{displaymath}
T_{max} = \mathrm{MJD}\ 55,994.6 \pm 0.4 + n \times 14.730 \pm 0.006\\
\end{displaymath}
where $T_{max}$ is the time of maximum flux.


The full amplitude of the modulation, defined as (maximum - minimum)/
mean flux, from the sine fit is approximately 70\%.
From the fundamental of the orbital peak in the power
spectrum we determine an orbital period of 3.73834 $\pm$ 0.00022 days,
while the second harmonic yields 3.73886 $\pm$  0.00014 days.
This is consistent with the period of 3.73886 +0.00028, -0.00140 days given by \citet{Levine2011}.
The BAT light curve of IGR J16418-4532 folded on the orbital period is also shown
in Figure \ref{fig:16418_fold} and this shows the presence of an eclipse.

\subsection{IGR J16479-4514} \label{section:16479}
IGR J16479-4514 is an SFXT with a rather short
orbital period of near 3.3 days with periods of 
3.3194 $\pm$ 0.0010 and 3.3193 $\pm$ 0.0005 days
determined by \citet{Jain2009} and \citet {Romano2009}, respectively, using Swift BAT
data in both cases. The folded light curve shows the presence of X-ray eclipses.
The mass donor has a spectral type of O8.5I \citep{Chaty2008,Rahoui2008}
or O9.5 Iab \citep{Nespoli2008}.
No X-ray pulsations have yet been reported.

\begin{figure}
\epsscale{0.8}
\includegraphics[angle=-90,scale=0.3]{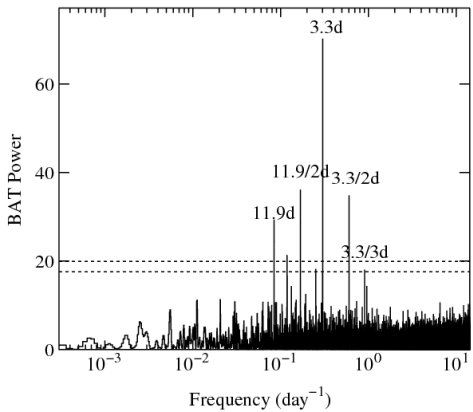} 
\caption{Power spectrum of the BAT light curve of IGR J16479-4514.}
\label{fig:16479_power}
\end{figure}

\begin{figure}
\epsscale{0.8}
\plotone{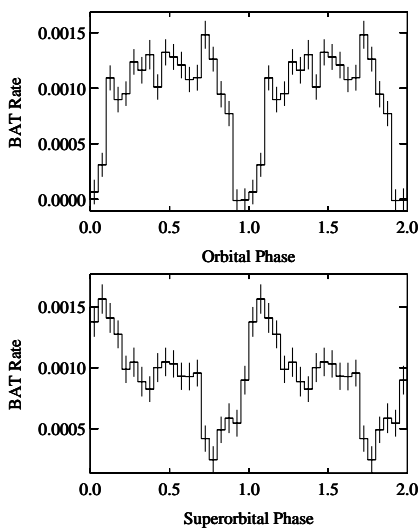}
\caption{Swift BAT light curve of IGR J16479-4514 folded on its orbital period {\mybf (top)}
and folded on its superorbital period {\mybf (bottom)}.
The period values are given
in Table \ref{table:sources}.
Phase zero for the orbital period is the center of the eclipse and
is MJD 54,547.05 \citep{Bozzo2009}.
Phase zero for the 
superorbital period corresponds to maximum flux and is MJD 55,996 (Section \ref{section:16479}).
}
\label{fig:16479_fold}
\end{figure}

The power spectrum of the BAT light curve (Figure \ref{fig:16479_power}) shows
modulation at the 3.32 day orbital period
and harmonics of this. From the fundamental we determine an orbital period
of 3.3199 $\pm$ 0.0005 days.
In addition to this, peaks are seen near 11.9 days
and its second harmonic.  
The FAP of the harmonic is 0.0006 while that of the fundamental is 0.05.
The second harmonic is stronger than the
fundamental and from this we derive a superorbital period of 11.880
$\pm$ 0.002 days.  The period determined from the fundamental is consistent
with this at 11.871 $\pm$ 0.005 days.
The BAT light curve of IGR J16479-4514 folded on the orbital and superorbital periods is shown
in Figure \ref{fig:16479_fold}. An eclipse is clearly seen in the light curve folded on
the orbital period.
The light curve folded on the superorbital period shows a
relatively sharp rise from minimum to maximum followed by a
plateau. The time of minimum flux is approximately MJD 55,993 $\pm$ 1.0
with maximum flux occurring approximately 0.25 in phase after this.
The full amplitude of the modulation is approximately 130\%.

\subsection{4U 1909+07 (X 1908+075)} \label{section:1909}
\citet{Wen2000} found a 4.400 $\pm$ 0.001 day orbital period for the X-ray binary 4U 1909+07
using \RXTE\ ASM observations.
X-ray pulsations with a period of 605 s were found with the \RXTE\ PCA
by \citet{Levine2004} and from a pulse arrival time analysis they
found the orbit to be circular with an orbital period of 4.4007 $\pm$ 0.0009 days
and derived
a mass function of 6.1 M\sun.
Although \citet{Levine2004} proposed that the primary might be a Wolf-Rayet star,
\citet{Morel2005} identified a likely near-IR candidate
which they proposed to be a late O-type supergiant.
\citet{Levine2004} found large orbital phase dependence of the X-ray absorption.
The orbital period was further refined with additional ASM observations to
4.4005 $\pm$ 0.0004 days by \citet{Wen2006}.

\begin{figure}
\epsscale{0.8}
\includegraphics[angle=-90,scale=0.3]{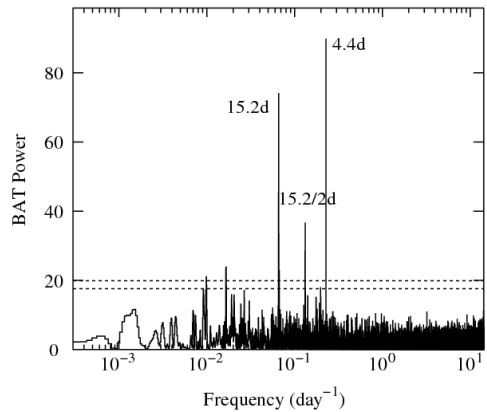}
\caption{Power spectrum of the BAT light curve of 4U 1909+07.}
\label{fig:1909_power}
\end{figure}

The power spectrum of the BAT light curve of 4U 1909+07 (Figure \ref{fig:1909_power}) shows strong
modulation at the orbital period and we derive a period of
4.4003 $\pm$ 0.0004 days. 
In addition, significant modulation at a
superorbital period near 15.2 days (FAP \sqig10$^{-5}$) and
the second harmonic of this are seen.  
Combining the detections at the fundamental
and second harmonic, we determine a period of 15.180 $\pm$ 0.003 days.


\begin{figure}
\epsscale{0.8}
\plotone{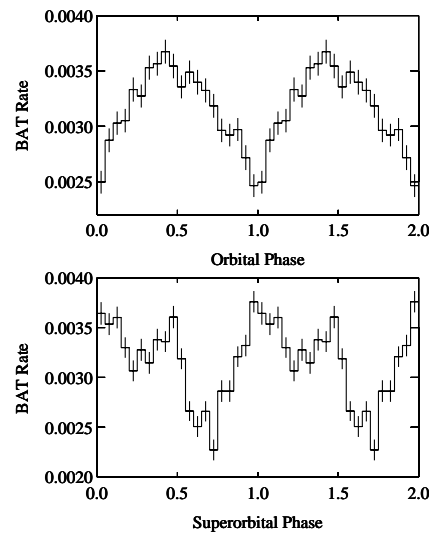}
\caption{Swift BAT light curve of 4U 1909+07 folded on its orbital period {\mybf (top)}
and folded on its superorbital period {\mybf (bottom)}.
The period values are given
in Table \ref{table:sources}.
Phase zero for the orbital period is the time of superior conjunction from
\citet{Levine2004} and is MJD 52,631.383. 
Phase zero for the superorbital period is the time of maximum flux
and is MJD 56,004 (Section \ref{section:1909}).
}
\label{fig:1909_fold}
\end{figure}

As expected from the presence of harmonics in the power spectrum, the
light curve folded on the superorbital period (Figure \ref{fig:1909_fold})
shows a multi-peaked
profile. The minimum is somewhat more clearly defined than the
maximum.  From an inspection of the folded light curve, the minimum
occurs at approximately MJD 55,999 $\pm$ 1.5. The time of maximum flux
occurs about 0.35 in phase after the minimum. The amplitude
of the modulation, defined as (maximum - minimum)/ mean flux
is approximately 50\%.
The BAT light curve folded on the orbital period is shown in Figure \ref{fig:1909_fold}.
This shows a quasi-sinusoidal modulation with no evidence for the presence of an eclipse.

\section{Properties of Selected Other sgHMXBs}

For comparison with the sgHMXB systems discussed above where superorbital modulation
is seen, we present here examples of systems where there is no strong superorbital modulation,
and two examples of systems which appear to be weak candidates for also possessing superorbital
modulation.

\subsection{Examples of Systems with Strong Orbital Modulation but Lacking Superorbital Modulation}

The Swift BAT set of light curves includes a number of other sgHMXBs.
However, the majority of these do not show evidence for superorbital modulation.
As examples we discuss here three systems. We choose ``IGR'' systems which are
typically rather hard sources and so suitable for study with the BAT. The examples
selected here all have very significant orbital modulations of their light curves
which have previously been reported.

\subsubsection{IGR J18027-2016 (= SAX J1802.7-2017)}
\begin{figure}
\epsscale{0.8}
\includegraphics[angle=-90,scale=0.3]{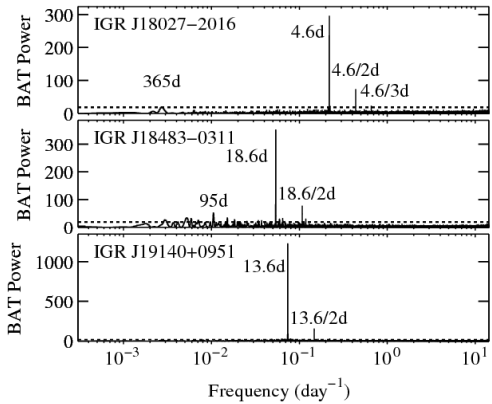}  
\caption{Power spectra of the BAT light curves of 
{\mybf IGR J18027-2016 (top), IGR J18483-0311 (middle), and
IGR J19140+0951 (bottom) The horizontal dashed lines indicate ``white noise'' 99.9\% and
99.99\% significance levels.}.
}
\label{fig:null_power}
\end{figure}

IGR J18027-2016 (= SAX J1802.7-2017) has a pulse period of 139.6s \citep{Augello2003}
and pulse arrival time analysis suggested a \sqig 4.6 day orbital period.
From a timing analysis \citet{Hill2005} refined this to
4.5696 $\pm$ 0.0009 days.
The spectral type of the mass donor has been proposed to be B1 Ib \citep{Torrejon2010}
and B0-B1 I \citep{Mason2011}, thus making it an sgHMXB.
The power spectrum of the BAT light curve of IGR J18027-2016 (Figure \ref{fig:null_power}, bottom panel) is
very flat with the exception of the orbital period and
its second and third harmonics, together with a small peak corresponding to a
period of one year.

\subsubsection{IGR J18483-0311}
{\mybf IGR J18483-0311 is an SFXT with an early B supergiant optical counterpart
\citep{Rahoui2008b}.}
Orbital modulation is seen at a period near 18.55 days
in
RXTE ASM \citep{Levine2011}, BAT \citep{Jain2009} and INTEGRAL observations \citep{Sguera2007}.
{\mybf This source} also has a 21 s pulse period \citep{Sguera2007}.
The power spectrum of the BAT light curve of IGR J18483-0311 (Figure \ref{fig:null_power}, middle panel) shows
strong modulation at the orbital period and the second harmonic of this.
The power spectrum exhibits somewhat larger ``noise'' at intermediate frequencies. 
A small non-statistically significant \sqig95 day bump is the third highest peak.

\subsubsection{IGR J19140+0951 (= IGR J19140+098)} 
IGR J19140+0951 (= IGR J19140+098) was discovered with INTEGRAL \citep{Hannikainen2004}
and a 13.558 $\pm$ 0.004  day period was found from RXTE ASM and Swift BAT observations \citep{Corbet2004}.
From infrared observations the optical counterpart was determined to be a
B0.5 supergiant \citep{Hannikainen2007}, later refined to B0.5 Ia
by \citet{Torrejon2010}. 
No pulsations have yet been reported from this source despite INTEGRAL and RXTE PCA observations \citep{Prat2008}.
The power spectrum of the BAT light curve of IGR J19140+0951 (Figure \ref{fig:null_power}, top panel) shows
an extremely flat power spectrum with the exception of strong peaks at the orbital period
and the second harmonic of this.

\subsection{Sources of Potential Superorbital Interest}
Although the presence of superorbital periods in sgHMXBs does not appear to be
ubiquitous, we can examine the power spectra of other wind-accretion HMXBs
for the possible presence of superorbital periods under the assumption that the
apparent correlation between orbital period and superorbital periods discussed in
Section \ref{section:correlation} is indeed
correct. This then yields a restricted frequency range to be searched for superorbital
modulation.

\subsubsection{1E 1145.1-6141} \label{section:1145}
The spectral type of the primary of 1E 1145.1-6141 was found to be 
B2 Iae by \citet{Hutchings1981} and \citet{Densham1982}.
The pulse period is \sqig297 s and pulse timing
enabled \citet{Ray2002} to determine a 14.365 $\pm$ 0.002 day orbital period with a modest
eccentricity of 0.20 $\pm$ 0.03.
\citet{Ray2002} report that no eclipse was seen.
No detection of orbital modulation of the X-ray flux from \RXTE\ ASM observations is reported
in the papers of 
\citet{Wen2006} and \citet{Levine2011}.
However, \citet{Corbet2007b} reported detection of the orbital period of
1E 1145.1-6141 in \Swift\ BAT data with the presence of flares at both
periastron and apastron. The presence of flares at 
apastron is also reported from \INTEGRAL\ observations by \citet{Ferrigno2008}.

\begin{figure}
\epsscale{0.8}
\includegraphics[angle=-90,scale=0.3]{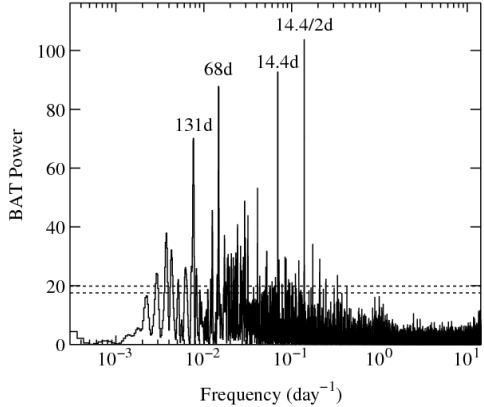} 
\caption{Power spectrum of the BAT light curve of 1E 1145.1-6141.}
\label{fig:1145_power}
\end{figure}

For 1E 1145.1-6141, the strongest peak in the power spectrum of the BAT
light curve (Figure \ref{fig:1145_power}) is at the second harmonic of the 14.4 day orbital period
and the second highest peak
is at the orbital period itself. 
The blind-search FAPs of the fundamental and second harmonic peaks would
be 0.1 and \sqig10$^{-5}$ respectively. The much lower significance of the
fundamental is due to the increase in continuum power at lower frequencies.
From the second harmonic we derive an orbital period of 14.365 $\pm$ 0.003 days,
which is the same as that derived by \citet{Ray2002} from pulse timing.
The peak at the fundamental yields a period of 14.373 $\pm$ 0.007 days, which is also
consistent, although with a somewhat larger uncertainty.

\begin{figure}
\epsscale{0.8}
\plotone{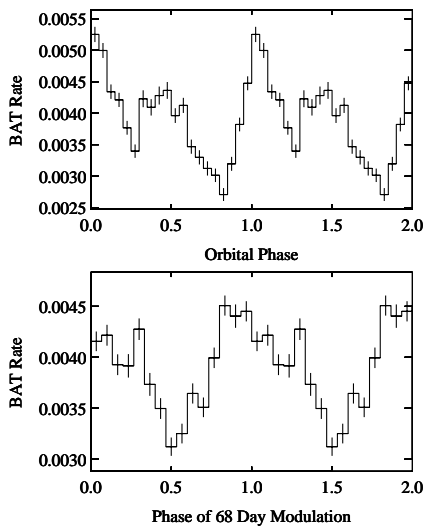}
\caption{Swift BAT light curve of 1E 1145.1-6141 folded on its orbital period 
{\mybf (top)}
and folded on a 67.8 day period {\mybf (bottom)}.
For the orbital period phase zero corresponds to the time of periastron
passage (MJD 51,008.1) determined by \citet{Ray2002}.
For the 68 day modulation, phase zero corresponds to the time of maximum
flux from a sine wave fit to the BAT light curve and is MJD 55,142.4 (Section \ref{section:1145}).
}
\label{fig:1145_fold}
\end{figure}

The BAT light curve folded on the orbital period (Figure \ref{fig:1145_fold})
shows a double-peaked profile with maxima at periastron and apastron
based on the ephemeris of \citet{Ray2002}. The third and the fourth highest
peaks in the power spectrum are at periods of 67.8 $\pm$ 0.2
(equivalent to 135.6 $\pm$ 0.4, if regarded as a second harmonic) and 131.4 $\pm$ 0.8 days.
The very low FAP of 0.2 of the \sqig68 day peak means that this is not a strong candidate 
for a superorbital period. However, the large
amount of variability in the light curve compared to the orbital modulation makes
this a potentially interesting system to continue to monitor.
From a sine wave fit to the BAT light curve, we derive an epoch of maximum flux 
for the 68 day modulation of MJD 55,142.4 $\pm$ 0.6.
The BAT light curve folded on the 68 day period is shown in Figure \ref{fig:1145_fold}.

\subsubsection{IGR J16393-4643 (= AX J16390.4-4642)} \label{section:16393}
The 910 s X-ray pulsar IGR J16393-4643 was reported by \citet{Thompson2006} to have
a 3.7 day orbital period from a pulse timing analysis, although other solutions with orbital periods of 
50.2 and 8.1 days could not be excluded.
\citet{Thompson2006} proposed, on the basis of their orbital parameters, that IGR J16393-4643 is a 
supergiant wind-accretion powered HMXB.
\citet{Nespoli2010} instead suggested that this is a symbiotic X-ray binary with a 50 day period. 
However, \Swift\ BAT and PCA Galactic plane scan
observations clearly showed the system to have a 4.2 day orbital period \citep{Corbet2010a}
which is consistent with an interpretation of the system as an sgHMXB.
The periods obtained from the BAT and PCA were 4.2368 $\pm$ 0.0007 and 4.2371 $\pm$ 0.0007 days
respectively.

\citet{Bodaghee2012} obtained a precise position for IGR J16393-4643 using a Chandra observation,
which excluded a previously proposed counterpart
that had led to the symbiotic classification by \citet{Nespoli2010}, and instead suggested
that the correct counterpart to IGR J16393-4643 might be a distant reddened star.

The 4.2 {\mybf day} orbital period of IGR J16393-4643 is similar 
to the 4.4 day orbital period of 4U 1909+07 which has a superorbital period of 15.2 days. 
If there is indeed a relationship between superorbital and orbital periods, as discussed in
Section \ref{section:correlation}, a superorbital period
of \sqig15 days would thus be predicted.

\begin{figure}
\epsscale{0.8}
\includegraphics[angle=-90,scale=0.3]{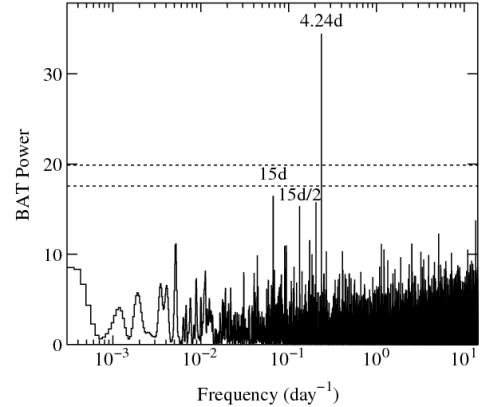}  
\caption{Power spectrum of the BAT light curve of IGR J16393-4643.}
\label{fig:16393_power}
\end{figure}

\begin{figure}
\epsscale{0.8}
\includegraphics[angle=-90,scale=0.3]{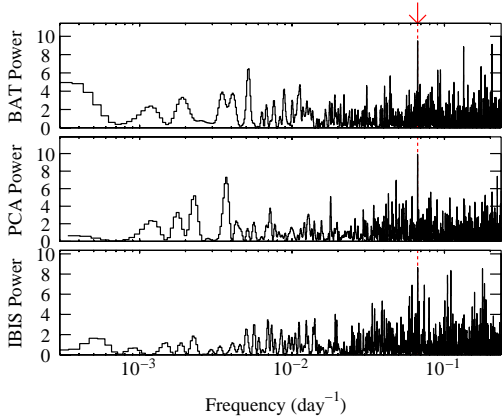}  
\caption{Power spectra of Swift BAT {\mybf (top), RXTE PCA Galactic Plane
scan (middle) and \INTEGRAL\ (bottom)} light curves of IGR J16393-4643
for periods longer than the orbital period.
The vertical dashed red lines and the arrow mark the possible superorbital period. 
}
\label{fig:16393_super_power}
\end{figure}

\begin{figure}
\epsscale{0.8}
\plotone{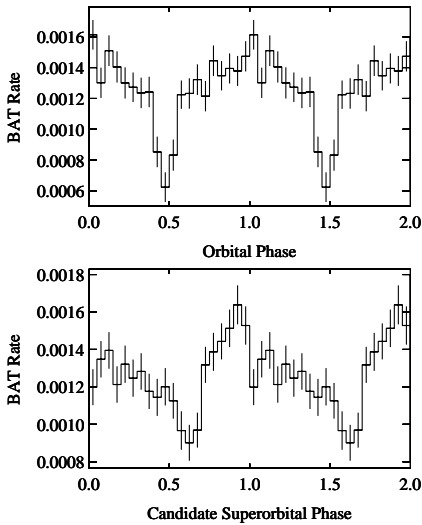}
\caption{Swift BAT light curve of IGR J16393-4643 folded on its orbital period {\mybf (top)}
and folded on its possible superorbital period {\mybf (bottom)}.
Period values are given in Table \ref{table:sources}.
Phase zero for the orbital period is time of maximum flux from \citet{Corbet2010a} (MJD 54,352.50) and
for the possible superorbital period it is MJD 55,092.6, the epoch of maximum flux from a sine
wave fit to the BAT light curve (Section \ref{section:16393})
}
\label{fig:16393_fold}
\end{figure}

The strongest peak in the power spectrum of the updated \Swift\ BAT light curve
(Figure \ref{fig:16393_power})
is at the orbital period with a value of 4.2380 $\pm$ 0.0005 days.
The second highest peak in the power spectrum 
(Figures \ref{fig:16393_power} and \ref{fig:16393_super_power}) is at a period  
near 15 days at 14.99 $\pm$ 0.01 days, and there is another peak at half this period.
The possible second harmonic is at a period of 7.485 $\pm$ 0.002 days, equivalent
to a fundamental period of 14.971 $\pm$ 0.005 days if regarded as a harmonic.
However, the ``blind search'' FAPs of both peaks, even restricting ourselves to
a search of periods longer than the orbital period, are very high at \sqig17\% and \sqig7\% for the 15 and 15/2 day
peaks respectively.
From a sine wave fit to the BAT light curve, with the period held fixed at 14.99 days,
we obtain an epoch of maximum flux of MJD 55,092.6 $\pm$ 0.4.
The BAT light curve of IGR J16393-4643 folded on the orbital period and the possible 14.99 day period
are shown in Figure \ref{fig:16393_fold}. The folded profile on the 14.99 day period suggests any modulation
may not be perfectly sinusoidal, with the maximum slightly preceding the value
predicted by the sine wave fit. The light curve folded on the orbital period suggests the presence
of an eclipse.

We therefore also investigated the PCA Galactic plane scans of this source obtained
between MJD 53163 and 55863 (2004-06-07 and 2011-10-29). This is 566 days longer than
the light curve used in \citet{Corbet2010a}.
The PCA data yield an orbital period of 4.2376 $\pm$ 0.0005 days, consistent with
our updated BAT result. In the power spectrum of the PCA scan observations (Figure \ref{fig:16393_super_power}), the largest
peak for periods longer than the orbital period 
is at 14.99 $\pm$ 0.01 days, consistent with the possible BAT period.

We next examined the \INTEGRAL\ light curve of IGR J16393-4643 that we obtained
from the Heavens service of the ISDC.
This covers a time range of MJD 52,650 to 55,856 (2003-01-11 to 2011-10-22) although
{\mybf with} only very sparse sampling. This limited sampling yields large amounts of artifacts
in the power spectrum of the light curve. For example, the 4.2 day orbital period is
not the strongest peak. However, examining the peak nearest the orbital period
yields a value of 4.2382 $\pm$ 0.0006, consistent with the periods derived from
the BAT and PCA observations. The \INTEGRAL\ power spectrum for periods
longer than the orbital period (Figure \ref{fig:16393_super_power}) also shows
a small peak near 15 days. This has a value of 14.98 $\pm$ 0.01 days, which is consistent
with the periods obtained from the BAT and PCA observations.

\begin{figure}
\epsscale{0.8}
\plotone{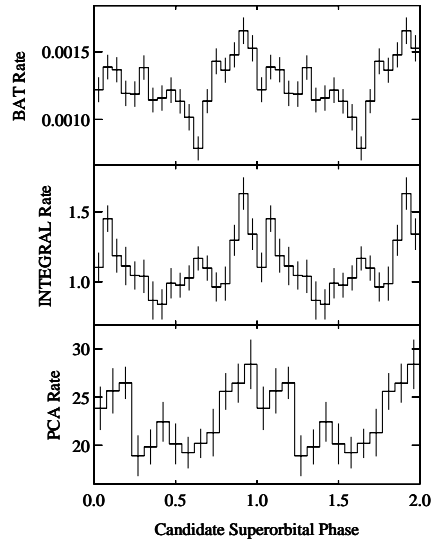}
\caption{Swift BAT {\mybf (top)}, INTEGRAL IBIS (middle)
and RXTE PCA scan {\mybf (bottom)} light curves of IGR J16393-4643 folded on its possible superorbital period.
Phase zero is MJD 55,092.6, the epoch of maximum flux from a sine
wave fit to the BAT light curve (Section \ref{section:16393})
}
\label{fig:16393_multi_super_fold}
\end{figure}

The BAT, PCA, and \INTEGRAL\ data folded on the possible superorbital
period derived from the BAT data are shown in Figure \ref{fig:16393_multi_super_fold}. 
The three light curves appear to have roughly coincident maxima. 
While the coincidence of the periods obtained from three separate instruments
is intriguing, additional data are required to confirm whether there truly is
a superorbital period in this system. {\mybf Such data may come from continued
monitoring with the BAT, or from future missions such as the proposed
Wide Field Monitor \citep[WFM;][]{Bozzo2013} on board the Large Observatory for X-ray Timing
\citep[LOFT;][]{Feroci2012}.}

\section{Discussion}

\subsection{Excluding Period Artifacts} \label{section:artifacts}

Since the presence of superorbital periods in sgHMXBs
is somewhat surprising, we consider whether they could be some type of artifact.
We note that they are not present in other types of systems, and there is no obvious way to create superorbital
modulation in only a subset of supergiant wind accretors.
Superorbital modulation in \2s\ and IGR J16493-4348 was previously seen in other detectors (\RXTE\ ASM and \RXTE\ PCA 
Galactic plane scan data respectively).
In several cases there are pulse arrival time orbits that show that the
orbital period really is the orbital period.
In addition, \citet{Drave2013b} report that \INTEGRAL/IBIS data confirm
the superorbital periods found for 4U 1909+07, IGR J16418-4532, and IGR J16479-4514.

The superorbital periods are rather prominent relative to the orbital
periods in the BAT energy range.
This appears to differ from results of lower-energy observations such as \RXTE\ ASM
observations of 4U 1909+07, where the orbital period is strongly detected,
but the superorbital period is not seen. Similarly, for \2s, although the superorbital period
was initially detected from \RXTE\ ASM data \citep{Farrell2006}, the superorbital
modulation has lower amplitude than the orbital modulation in the ASM energy band. In contrast, the
superorbital modulation of \2s\ is stronger than the orbital modulation in the
BAT observations. One reason for this is likely to be the lower fractional modulation of
the X-ray flux on the orbital period in
the BAT energy range for non-eclipsing systems. For these types of systems
a large component of the orbital modulation seen with lower-energy instruments
is due to the changing absorption as
the neutron star orbits its companion, which the BAT is relatively insensitive to.

\subsection{Coherence}
Superorbital modulation from Roche-lobe overflow systems is not
necessarily coherent. For example, there is considerable variation in
the superorbital periods of 
SMC X-1 \citep{Coe2013,Wojdowski1998} and Her X-1 \citep{Leahy2010}, although
not in LMC X-4 \citep{Hung2010}.
The sampling of the BAT light curves is very variable. This potentially makes
it more difficult to calculate the true resolution of the power spectra.
Therefore, in order to investigate the coherency of the superorbital modulation, we
compared the widths of the superorbital peaks to those of the orbital peaks.
This uses the assumption that the orbital modulation should be essentially periodic.
We fitted Gaussian functions to the superorbital and orbital peaks in the power spectra
in the five systems
for which superorbital modulation is definitely observed and determined the
following ratios of superorbital to orbital peak widths:
\2s, 1.05; IGR J16493-4348, 1.16; IGR J16418-4532, 0.92; IGR J16479-4514, 0.98;
4U 1909+07, 0.92. 
We therefore conclude that the superorbital modulations have very high coherence.
For the two low-significance candidates we obtain ratios
of superorbital to orbital peak widths of: IGR J16393-4643, 0.66; 1E 1145.1-6141, 0.89.

\subsection{Relationship Between Superorbital and Orbital Periods}\label{section:correlation}
\begin{figure}
\epsscale{0.8}
\includegraphics[angle=-90,scale=0.3]{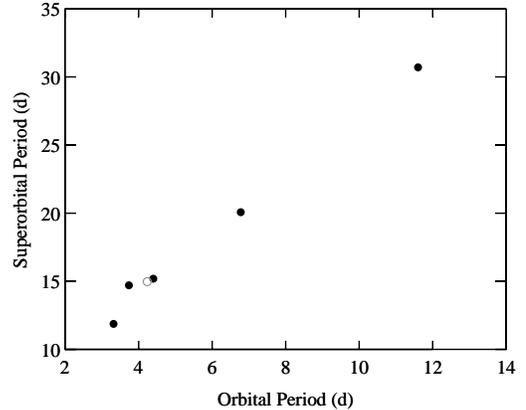}
\caption{Superorbital period plotted against orbital period for
the wind-accretion HMXBs discussed in the text. Statistical uncertainties
on period measurements are smaller than symbol sizes.
The filled symbols show definite superorbital period detections.
The gray open symbol marks
the modulation seen in IGR J16393-4643 which is
not yet considered to be a definite detection of a superorbital period.
}
\label{fig:superplot}
\end{figure}

System parameters are summarized in Table 1, and
in Figure \ref{fig:superplot} we plot superorbital period against orbital period.
For the five systems with definite superorbital modulation, the linear correlation
coefficient between P$_{super}$ and P$_{orb}$ is 0.996
and the associated probability of obtaining this level of correlation
from a random data set is 0.03\%. 
Because this possible correlation comes from such a small number of systems,
a determination whether this possible dependence of superorbital period on orbital period
is correct requires the candidate superorbital period in IGR J16393-4643 to be investigated
with additional data, and further superorbital periods
must be found in other systems. 
For the five systems the best linear fit for superorbital periods vs. orbital period
has parameters of:
\begin{displaymath}
P_{super} = 2.2 \pm 0.1 \times P_{orb} + 5.6 \pm 0.8\ \mathrm{days}
\end{displaymath}

\begin{figure}
\epsscale{0.8}
\includegraphics[angle=-90,scale=0.3]{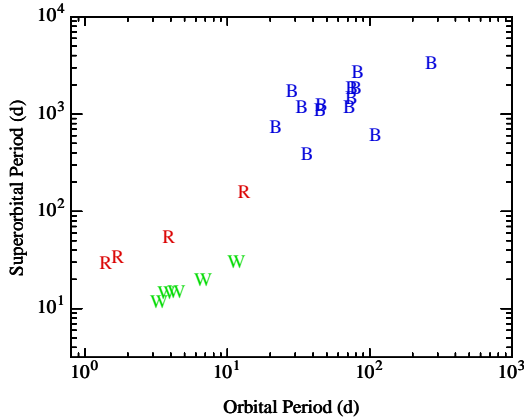}
\caption{Superorbital period plotted against orbital period for
the a variety of HMXBs including both neutron star and black hole systems.
``R'' indicates Roche-lobe overflow systems, ``W'' are the five wind-accretion systems
discussed in this paper, and ``B'' shows Be star system parameters taken
from \citet{Raj2011}.
The sources included as Roche-lobe overflow systems are the high-mass neutron star
systems LMC X-4 and SMC X-1, the intermediate-mass neutron star system Her X-1,
and the black hole candidate SS 433. 
}
\label{fig:duperplot}
\end{figure}

For comparison, in Figure \ref{fig:duperplot} we plot superorbital period against
orbital period for a wide variety of systems. These include these Roche-lobe overflow powered
systems: LMC X-4 (neutron star HMXB, P$_{orb}$ = 1.4 days, P$_{super}$ = 30.3 days), Her X-1 (neutron star intermediate-mass system,
P$_{orb}$ =1.7 days,  P$_{super}$ = 35 days), SMC X-1 (neutron star HMXB, P$_{orb}$ = 3.89 days, P$_{super}$ = 56 days), 
and SS 433 (black hole candidate microquasar, P$_{orb}$ = 13.1 days, P$_{super}$ = 162.5 days).
These parameters are taken from \citet{Kotze2012}.
Be star systems are also shown with their parameters taken from \citet{Raj2011}.
For the Be star systems the superorbital modulation periods may be quasi-periodic rather than
strictly periodic. 
It the mechanisms driving superorbital
modulation differ between different types of object, then the different types
of system could be located in different regions of this diagram.
We note that the sgHMXB superorbital periods are rather
short relative to their orbital periods, compared to other types of systems.
This is suggestive that, as expected, a different driving mechanism may be at work in the
sgHMXBs superorbital modulation compared to the other types of system.

\subsection{Possible Mechanisms Driving Superorbital Modulation}

We note that \citet{Farrell2008}'s extensive \RXTE\ PCA observations of \2s\ showed
that there were changes in absorption over the orbital period of this
system, but not on the superorbital period. \citet{Farrell2008} therefore concluded
that the superorbital modulation was related to variability in the
mass accretion rate caused by an unknown mechanism. This is consistent with
the stronger detection of superorbital periods with the BAT compared to
orbital periods for non-eclipsing systems than is the case with the RXTE ASM (Section \ref{section:artifacts}).
The ASM is more sensitive to changes in absorption over the orbital period. However, the BAT
has overall better sensitivity for changes in the flux of highly-absorbed systems.

If mass-transfer rate variations are the cause of periodic superorbital modulation,
then this suggests that the wind from the primary star could itself be modulated
in some way by a mechanism related to the length of the orbital period.
However, any satisfactory model must be able to account for the lack of
superorbital variability in many systems. Thus, the modulation must be related
to some parameter independent of inclination angle and whether the system
is an SFXT or classical supergiant system. For example, an offset between
the orbital plane and the rotation axis of the primary star, or a small
orbital eccentricity, might satisfy
such a requirement.
The light curves folded on the superorbital periods exhibit a variety of 
morphologies, and so any model for the modulation must be able to account for this.
We note the work of \citet{Koenigsberger2003} and \citet{Moreno2005} indicates that
oscillations can be induced in non-synchronously rotating stars in binary
systems on periods longer than the orbital period.
As discussed by \citet{Koenigsberger2006}, such oscillations result in changes in the mass-loss rate of the
primary star which would cause correlated changes in the X-ray luminosity. 
This model may also be consistent with the lack of superorbital modulation in
all systems if only a fraction of primary stars are rotating non-synchronously.
However, as noted by \citet{Farrell2008}, the \citet{Koenigsberger2006} model applies
to circular orbits and \2s\ has a modest eccentricity of 0.2.
In addition, the high coherency of the superorbital modulations seems difficult
to account for with oscillations of the primary star unless there is
some way to keep strict phase stability.

In principle, the presence of a third object in the systems might drive superorbital
modulation. The modulation from a third body would also naturally account for the coherency of
the superorbital modulation.
However, the apparent correlation between superorbital and orbital periods
would, if confirmed, place stringent constraints on how such multi-star systems might
be formed. In addition, typically such triple body models for X-ray sources involve
hierarchical systems with a distant third object \citep[e.g.][]{Mazeh1979}.

\section{Conclusion}

Observations with the \Swift\ BAT have shown the presence of superorbital periods
in three additional sgHMXBs for a total of five definite such systems.
The superorbital modulations have a variety of morphologies, ranging from
approximately sinusoidal to multi-peaked profiles.
However, superorbital modulation is not a ubiquitous property of sgHMXBs.
With this limited set of data, a possible dependence of superorbital period
on orbital period is suggested.
The mechanism(s) driving such superorbital modulation remain unclear.
However, possible models based on oscillations in the primary star driven
by non-synchronous rotation, and three-body systems deserve further investigation.
Continued monitoring of sgHMXBs in hard X-rays both with additional \Swift\ BAT
data and also potential new missions with high-energy X-ray sensitivity 
such as the {\mybf LOFT WFM} 
may reveal additional sources with superorbital periodicities.

\acknowledgements
{\mybf We thank an anonymous referee for useful comments.}
This paper used \Swift/BAT transient monitor results provided by the \Swift/BAT team.
The Swift/BAT transient monitor and H. A. K. are supported by NASA under Swift Guest Observer
grants NNX09AU85G, NNX12AD32G, NNX12AE57G and NNX13AC75G.

\pagebreak



\begin{deluxetable}{lccccccc}
\tablewidth{6.5in}

\tablecaption{Wind-Accretion sgHMXBs with Periodic Superorbital Modulation}

\tablehead{
\colhead{Name} & \colhead{P$_{orb}$} & \colhead{P$_{super}$} &
\colhead{P$_{super}$/P$_{orb}$} &
\colhead{P$_{spin}$} & \colhead{Spectral Type} & \colhead{Eccentricity} &\colhead{SFXT?}\\
\colhead{} & \colhead{(days)} & \colhead{(days)} &
\colhead{} &
\colhead{(s)} & \colhead{} & \colhead{} \\
}

\startdata
IGR J16479-4514 &  3.3199 $\pm$ 0.0005 &  11.880 $\pm$ 0.002 days & 3.58 & ? & O8.5I/O9.5 Iab & ? & Y \\
IGR J16418-4532 &  3.7389 $\pm$ 0.0001 &  14.730 $\pm$ 0.006 &  3.94 & 1212 & {\mybf O8.5} & ? & Y \\
4U 1909+07 & 4.4003 $\pm$ 0.0004  &  15.180 $\pm$ 0.003 & 3.45 & 605 & late O & 0.02 $\pm$ 0.04 & N \\
IGR J16493-4348  &  6.782 $\pm$ 0.001 & 20.07 $\pm$ 0.01 & 2.96 & 1093 & B0.5−1 Ia-Ib & ? & N \\
2S 0114+650  &  11.591  $\pm$ 0.003 & 30.76 $\pm$ 0.03 & 2.65  & \sqig9700 & B1 Ia & 0.18 $\pm$ 0.05 & N\\
\tableline
IGR J16393-4643 & 4.2380 $\pm$ 0.0005 & (14.99 $\pm$ 0.01) & (3.54) &  910 & ? & ? & N \\
\enddata
\label{table:period_fits}
\tablecomments{The superorbital period for the system below the line is a candidate
and not a definite detection. The references for system parameters are given in the individual
sections on each source. The orbital and superorbital periods and their errors are derived from the BAT light curves.
For some systems additional determinations of periods may be available from other work, as given in the individual
source sections.
}
\label{table:sources}
\end{deluxetable}

\end{document}